\documentclass[]{revtex4-1}
\usepackage{amsfonts,amsmath,amssymb,graphicx,subfigure}

\begin{document}

\title{Is the immune network a complex network?}

\author{Hallan Souza-e-Silva}
\email{hallan.silva@ufv.br, Permanent Address: Departamento de F\'isica, Centro de Ci\^encias Exatas, Universidade Federal de Vi\c cosa, Vi\c cosa, MG, Brasil}

\author{Rita Maria Zorzenon dos Santos}
\email{zorzenon@df.ufpe.br, corresponding author}
\affiliation{Departamento de F\'isica, Centro de Ci\^encias Exatas e da Natureza, Universidade Federal de Pernambuco, Recife, PE, Brasil}

\begin{abstract}

Some years ago a cellular automata model was proposed to describe the evolution of the immune repertoire of B cells and antibodies based on Jerne's immune network theory and shape-space formalism. Interactions among different B-cell clones, which may be found in low, intermediate and high concentrations, occur depending on the complementarities of their characteristic proteins and are regulated by activation and suppression mechanisms. Depending on the region of the parameter space, the model exhibits either stable (ordered) or chaotic behavior, but it is in the transition region of the parameter space between both regimes where we obtain the complex behavior of a self-regulated network, much like Jerne’s idea of the immune network. In this region, the network maintains the memory of the large perturbations, which simulate antigen presentations and reproduce immunization and ageing experiments performed with mice. 
Here we investigate if the networks generated by this model in the different regimes can be classified as complex networks. We have found that in the chaotic regime the network has random characteristics with large, constant values of clustering coefficients, while in the ordered phase, the degree distribution of the network is exponential and the clustering coefficient exhibits power law behavior. In the transition region we observed a mixed behavior (random-like and exponential) of the degree distribution as opposed to the scale-free behavior reported for other biological networks. Randomness and low connectivity in the active sites allow for rapid changes in the connectivity distribution of the immune network in order to include and/or discard information and generate a dynamic memory. However it is the availability of the low concentration nodes to change rapidly without driving the system to pathological states that allow the generation of dynamic memory and consequently a reproduction of immune system behavior in mice.  Although the overall behavior of degree correlation is positive, there is an interplay between assortative and disassortative mixing in the stable and transition regions regulated by a threshold value of the node degree, which achieves a maximum value on the transition region and becomes totally assortative in the chaotic regime.

\end{abstract}


\maketitle

\section{Introduction}

The main task of the immune system is to protect the integrity and identity of the body against any harm.  The main cells of the immune system are macrophages and lymphocytes, the former being mostly responsible for innate immune responses and lymphocytes, which are responsible for adaptive or cell-mediated immune responses. 

All immune cells carry a large number of molecular receptors (proteins) on the surface and the immune system works based on pattern recognition. There are two main classes of lymphocytes: T and B cells. While T cells are involved in signaling and functional activities for the majority of the pathogens, B cells are mainly responsible for the production of antibodies, which in general, function as markers for the pathogens to be phagocyted and improve the efficiency of the immune response. The antibodies produced by any B-cell population are copies of its molecular receptors and according to the clonal selection theory \cite{Janeway} proposed by Burnet in 1959, the antigen (or its binding sites), by pattern recognition chooses the B-cell clones (population of B cell and antibodies) that will proliferate. 

According to estimates lymphocytes carry the order of $10^{5}$ molecular receptors on its surface and the human immune system is able to express the order of $10^{11}$ different receptors during its lifetime. Such large numbers allow for the recognition of any antigen presented to the immune system and for the completeness of the immune repertoire. If the repertoire is complete we should expect the elements of the immune system to recognize and be recognized by other elements; the same mechanism of recognition should also work for both antibody-antigen and antibody-antibody reactions. In 1974, Jerne \cite{Jerne}, taking these ideas into account, suggested that when the antigen is presented to the organism it will activate a set of B-cell clones and the production of specific antibodies (complementary to the antigen binding sites) that in turn would activate the receptors of other clones, and so on. Due to the interplay of activation and suppression mechanisms, the reaction chain would be finite, preventing the percolation of information through the entire system. This kind of dynamics generates a multi-connected network of cell populations that regulates the immune response, and at any time reflects the dynamic memory of the system regarding its previous history in terms of antigen presentation. In other words, the immune response to different pathogens (virus, bacteria, etc) is regulated by dynamics involving the complementary molecular receptors of the different B-cell clones that create a dynamic memory, which includes new and already existing information concerning the previous immunization process. This memory allows for a fast response to new presentations of previously seen antigens. 

Since its proposal, little evidence has been put forward to support the existence of the immune network theory \cite{Coutinho89, Holmberg-Forsgren, Lundkvist-Holmberg}, but research has suggested that if the network exists, then only $20\%$ of the lymphocytes will be activated, while the rest of the clones will form a pool of immunocompetent lymphocytes, which are able to recognize any antigen.

In 1992, Stauffer and Weisbush \cite{physA18042} proposed a cellular automata model to describe the immune network based on Jerne's ideas, shape-space formalism and a previous model  introduced by De Boer and Perelson \cite{DeBoer-Segel-Perelson}. This discrete model was later modified by Zorzenon dos Santos and Bernardes \cite{physa2191} and hereafter will be referred to as the BSP model. Using shape-space formalism, the model allows for the simulation of the large immune repertoire and the complementary interactions between B-cell clones. Each B-cell population is associated to a three-state automatum representing low, intermediate and high concentrations, and the interactions amongst the different populations are described by activation and suppression mechanisms. The dynamics of the model leads to stable (ordered) or chaotic behaviors that correspond to pathological rather than regular behaviors of the immune system.  However, in the transition region between the two regimes \cite{ physA-93} during very long periods, the model describes aggregation-disaggregation dynamics with clusters splitting and fusing throughout time, as a multi-connected network of populations \cite{jtb186173}. This network exhibits self-regulation and despite changes within the active populations, only 10-20\% of the populations remain active. This multi-connected network subjected to multiple perturbations was used with mice to simulate and reproduce the behavior of their immune systems under multiple antigen presentations and the effects of ageing on generating these responses \cite{prl813034}.

The dynamic behavior of the BSP model on the transition region was further investigated regarding different aspects.  It has especially been observed that the memory of the system is dynamically allocated in order to incorporate new information regarding the history of the system without losing previously acquired important information \cite{Zorzenon-Copelli03}. As the system becomes older, so less information is incorporated, which is a behavior associated to the loss of plasticity, as observed in other physical systems \cite {epjb34119}. 

Although there have been many studies on the parameter space and dynamics of the BSP model, it has not yet been characterized from the viewpoint of a complex network. Since it reproduces the behavior of real immune systems it is interesting to investigate its topological properties using graph theory formalism (see \cite{rmp7447} and references therein) adopted in the characterization of complex networks found in nature \cite{nature41141, rmp7447}.  From previous studies on the dynamic behavior of the immune network it becomes clear that the regulation of the dynamics leading to stable, complex or chaotic behaviors emerges from the interplay between subnetworks of populations with low, intermediate and high activations. Our aim in this work is to investigate whether the networks generated by the BSP model in the different regimes (stable, transitional and chaotic) may be classified as complex networks.  Therefore, we have focused on studying and characterizing the properties of these subnetworks for each phase with regard to degree distribution, the behavior of clustering coefficients, the existence of hierarchical structures and the correlations within the neighborhood (assortative or disassortative mixing behaviors) of the sites.

In what follows, in section \ref{model}, we introduce the SW cellular automata model proposed to describe the immune network, in section \ref{complexnet} we undertake a short review of the main concepts involved in the study of the above-mentioned properties, in section \ref{results} we present and discuss the results obtained and in section \ref{discussion} we make our final remarks.

\section{The cellular automata model} 
\label{model}

In the model introduced by Stauffer and Weisbuch \cite{physA18042} shape-space formalism is adopted to describe the repertoire of B-cell clones and the interaction between different clones \cite{jtb81645}. Each clone is represented by its molecular receptor, which corresponds to a point in a d-dimensional space. Thus, each receptor is characterized by $d$ properties corresponding to different characteristics, as for instance, the number of nucleotides, charge, hydrophobicity, etc. The nearest clone neighbors differ only by one property and the lock-key interactions occurring among populations with complementary molecular receptors is described by a non-local rule, where an on-site clone $\vec{r}$ is influenced by those located at $-\vec{r}$  and its nearest neighbors (representing slightly defective interactions). According to estimates, if the shape-space notion is relevant in a continuous mathematical approach, we obtain $d \geq 5$ \cite{jtb81645} but if it is relevant in a discrete approach, then it is $d \geq 2$ \cite{jtb186173}.
The population associated to each molecular receptor $\vec{r}$ is represented by a three-state automaton describing its concentration at any given time: low ($B(\vec{r}, t) = 0$), intermediate ($B(\vec{r}, t) = 1$) and high ($B(\vec{r}, t) = 2$).   The influence on the population of site $\vec{r}$ caused by its complementary populations is described by the field $h(\vec{r}, t)$:

\begin{equation}
h(\vec{r}, t)= \sum_{\vec{r}'  \epsilon -(\vec{r}+\delta \vec{r})} B(\vec{r}, t)\\
\end{equation}
where for each $\vec{r}$ the sum runs over the complementary shape $-\vec{r}$ and its nearest neighbors. Due to the finite number of B-cell population states, the maximum value of the field $h(\vec{r},t)$ is $h_{max} = 2(2d + 1)$. The rules describing the changes of B-cell populations due to their interactions with other populations are based on an activation window, which is inspired by a log-bell-shaped proliferation function associated to the receptor cross-linking involved in B-cell activation \cite{physA18042, physa2191, jtb186173}. In this window there is a minimum field necessary to activate the proliferation of the receptor populations ($\theta_1$), but there is also an upper limit for activation since for high doses of activation (greater than $\theta_2$) the proliferation is suppressed. The updating rules are summarized as follow:

\begin{equation}
B(\vec{r},t+1)= \left \{ \begin{array}{cc}\\
B(\vec{r},t) +1 & \mbox{if} \quad \theta_1\le h_i(t)\le\theta_2\\
B(\vec{r},t) -1    & \mbox{otherwise}\\
\end{array}\\
\right.
\end{equation}
 but no change is made if it would lead to $B = -1$ or $B = 3$. At each time step $t$ we define the densities of sites in state $i$ as $B_{i} (t)$, where $i $= 0, 1 and 2 correspond to low, intermediate and high concentrations, respectively.

The initial configurations are randomly generated depending on the parameter $x$ that controls the initial concentrations: $ B_{1}(0)$ = $B_{2}(0)$ =$ x/2 $, while the remaining $L^{d}(1-x)$ sites are initiated with low concentrations.

This model exhibits stable and chaotic regimes for $d \geq 2$ \cite{physa2191}, depending on the activation threshold ($\theta_1$) and the width of the activation window ($\theta_2-\theta_1$). These different dynamic behaviors are separated by a transition region where the model exhibits complex behavior during cycles of very long periods \cite{physA-93} in which clusters of different B-cell populations split and fuse \cite{jtb186173} and the activated populations behave like a multi-connected network. In 1998 \cite{prl813034} this multi-connected network was used to reproduce the results of experiments performed with mice and showed refractory behavior under multiple antigen presentations. Antigen presentation is simulated by the introduction of perturbations flipping the state of low concentration populations to those of high concentration, and computing the changes on the multi-connected network with respect to the initial state before the introduction of the perturbation. The same study has shown that the aging effects observed in the immune responses of mice can be explained by the loss of plasticity and the ability to bring about new changes in order to include new information in the system, i.e., the older the system, the more rapidly it saturates and the less intense is its response. It has also been found \cite{Bernardes-Zorzenon01} that there is a characteristic cluster size associated to the loss of plasticity and a power law distribution for the permanence times (the time interval that each population remains activated or belongs to the multi-connected network). As expected, the absence of scale indicates that there is no typical permanence time, a feature that may be understood since the memory of the system is allocated dynamically at each time step depending on the interactions among the populations. Since the aging effects observed in the multi-connected network have similarities with physical glassy systems, auto-correlation functions were obtained \cite{epjb34119}.  As previously mentioned, the system with no perturbation is driven to a long-limit cycle-attractor after a long transient time. When subjected to small, random perturbations however, the very notion of a transient becomes fuzzy and the results in Ref. \cite{epjb34119} show that the system is deflected from its attractor by small perturbations after $10^3-10^4$ time steps. Therefore, small perturbations will cause the system to change attractors from time to time due to their cumulative effects, and is reflected in the decreasing auto-correlation functions. Large perturbations would not accelerate the de-correlation process. In fact, large perturbations lead to a much weaker (slower) de-correlation, since they are always produced on the same sites in shape-space to simulate antigen presentations. Small perturbations can be more easily absorbed by the system than larger ones since they involve only local changes. The various studies performed until now \cite{jtb186173, epjb34119, Zorzenon-Copelli03} to understand the dynamics of the multi-connected network have suggested that different behaviors (stable, complex and chaotic) emerge from the interplay between the subnetworks of low, intermediate and high concentrations due to the activation of suppression mechanisms. In this work, we have investigated the properties of these inter-dependent sub networks from the viewpoint of complex networks using graph theory formalism.

\section{The complex network properties investigated}
\label{complexnet}

In what follows we will consider the multi-connected network and the subnetworks as graphs composed of nodes and links. The key quantity on the study of the topological properties of graphs and complex networks is the degree distribution of the nodes.  The connectivity or degree associated to the node $i$ is defined by the number $K_i$ of links that connects node $i$ to other nodes on the graph. In other words,

\begin{equation}
K_i=\sum_{j} a_{ij}
\end{equation}
 where the sum runs over all nodes $j$ not equal to $i$ and $a_{ij}$ corresponds to connectivity or adjacent matrix elements. If there is a link between $i$ and $j$,  $a_{ij}=1$, otherwise $a_{ij}=0$. The degree distribution obtained for the network indicate the topological class of universality (regular, random or scale-free) to which the complex network belongs and whether there is an existence hidden variables that may deviate its behavior from one of the main universality classes.

Due to the complementary interaction that regulates the interactions between the B-cell populations, another quantity of interest in the study of the multi-connected network is the clustering coefficient of a given node, which is a measure of the degree to which nodes in a graph tend to cluster together. The clustering coefficient of a node $i$ ($C_i$) is defined as:

\begin{equation}
C_i = 2\frac{NC}{K_{i}(K_{i}-1)}
\end{equation}
 where $NC$ is the number of connections between the neighbors of site $i$. The average of all nodes gives the network clustering coefficient $C$. While $C_i$ is a local property, $C$ is global. The global clustering coefficient gives an overall indication of the clustering in the network, whereas the local coefficient describes the embedding of single nodes. The behavior of C(K) as a function of K may indicate the presence of hierarchical structures in the network when it exhibits power-law behavior \cite{pre67026112}. Hierarchical structures go beyond simple clustering by including the simultaneous organization of all scales in a network. Such structures are represented by trees or dendrograms, in which closely related pairs of vertices (nearest neighbors) have the lowest common ancestors than more distantly related pairs. In the case of the immune network we should expect to find some hierarchical structures, since it is built through complementary interactions using shape-space formalism, where similar shapes should have common ancestors.

To complete the characterization of the subnetworks we investigate the existence of assortative or disassortative mixing on the different subnetworks, a measurement that will define if the highly-connected nodes in each subnetwork tend to have more connections with other highly-connected nodes or with nodes with a lower number of connections, respectively. In order to investigate the assortativity of the network we calculate the degree-correlation of the complex network ($K_{nn}$) \cite{knn_book}. When there is a tendency of the nodes to connect to other nodes with a similar degree, the correlation is assortative, while when there is a prevalence of links between nodes with dissimilar degrees, the correlations are disassortative. The average degree between nearest neighbors $K_{nn}$  \cite{knn_book} is defined as:

\begin{equation}
K_{nn}(K) = \frac{1}{N_K} \sum_{i/K_{i=k}} K_{nn}(i)
\end{equation}

The sum runs over all nodes with degree $K$, $N_K$ is the number of nodes of degree $K$ and $K_{nn}(i)$ is the average nearest neighbor’s degree of vertex $i$:

\begin{equation}
K_{nn}(i) = \frac{1}{K_i} \sum_{j}K_j
\end{equation}
where the sum runs over all nearest neighbors of node $i$. When $K_{nn}$ is the constant, the degrees of neighboring nodes are uncorrelated.

Another quantity of interest is the mean shortest path ($L$) that corresponds to the average number of connections between any two nodes $i$ and $j$ calculated over all pairs of nodes. For regular hypercubic lattices in d-dimension $L\sim N^{1/d}$ and in the case of random graphs $L$ grows logarithmically with the number N of nodes ($L\sim log N$. The small world effect corresponds to the case in which any pair of nodes is connected by the shortest path distance \cite{knn_book}.

By investigating these properties it was possible to better understand the dynamics of the immune network from a topological viewpoint and the interdependence of the three subnetworks.

\section{Results}
\label{results}

All the results reported in this section were obtained using the following parameters: $d=3$, $L=50$, $N=L^d$ sites, $\theta_1=1/3\times h_{max}$ and $\theta_2=2/3\times h_{max}$, where $h_{max}=2(2d+1)$ and they are representative of the behaviors obtained on the different regions analyzed. Thus, when varying the parameters we should expect variations in the localization and size of these regions in the parameter space, but the overall behavior of the quantities analyzed here in the different regions would be the same. For the sake of clarity we will refer to the subpopulations or subnetworks of different states of activation as: $B_0$ for low, $B_1$ for intermediate and $B_2$ for high. By subnetwork we mean the network formed by all nodes in the same state. In the present study, for the maximum number of neighbors of a given node $d=3$ is 24 and by neighbors of node $i$ we considered its nearest and next-nearest neighbors. Despite the fact that the dynamics of the model is based on complementary interactions and involve mirror images, the neighbors of a given site in a subnetwork correspond to the populations in the same state belonging to its neighborhood, as defined above. The average values used for the majority of graphs as well as the distributions, over 1000 samples were obtained, discarding the 1000 initial time steps, in order to guarantee that the samples correspond to thermalized states on the different regimes.

\begin{figure}[hbt]
    \centering
    \includegraphics[scale=0.33,angle=0]{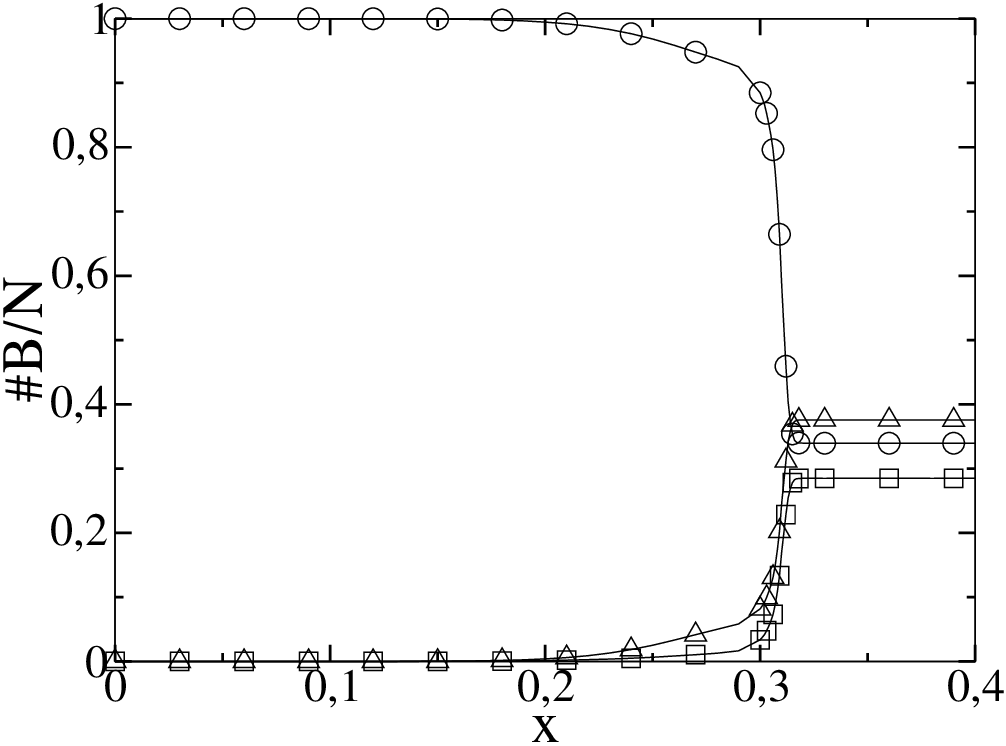}
    \caption{ The average population densities as a function of the parameter control of initial configurations ($x$). The symbols correspond to:  ($\bigcirc$) $B_0$; ($\square$) $B_1$ and ($\triangle$) $B_2$ concentrations. There is a remarkable change in the behavior of the densities for $x\sim 30$ separating the stable and chaotic phases.}
   \label{density}
\end{figure}

The transition region between stable to chaotic behaviors for this set of parameters is located in the vicinities of $x \sim 30$:  the stable phase corresponded to the region of $x<30$, and the chaotic associated to the region of $x>30$. Figure \ref{density} shows that the stable phase is dominated  by populations in low concentration ($B=0$), while in the chaotic phase the density of the three types of populations are of the same order of magnitude and therefore, $2/3$ of the nodes are activated ($B=1$ and $B=2$) corresponding to a pathological state as previously suggested \cite{physA-93}. All values shown in Figure \ref{density} correspond to the values of the densities calculated after thermalization. 

\begin{figure}[hbt]
   \centering
   \subfigure[]{\label{degree_all_x020}}{\includegraphics[scale=0.33,angle=0]{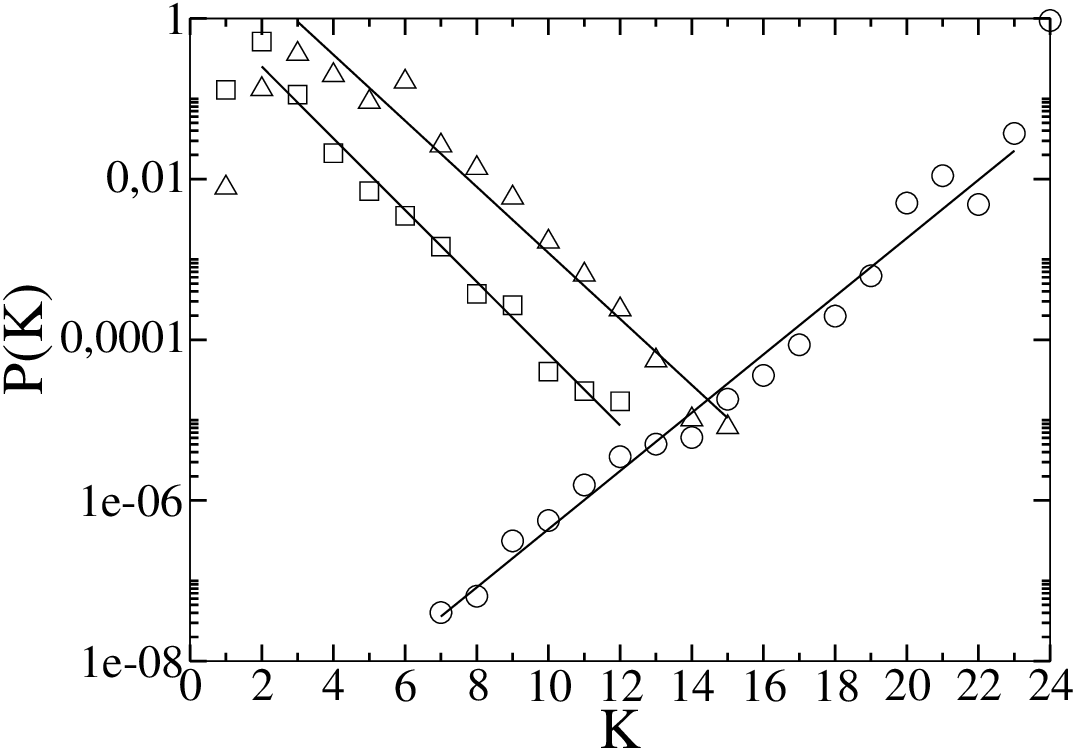}}
   \hspace*{0.25cm}\subfigure[]{\label{degree_all_x040}}{\includegraphics[scale=0.33,angle=0]{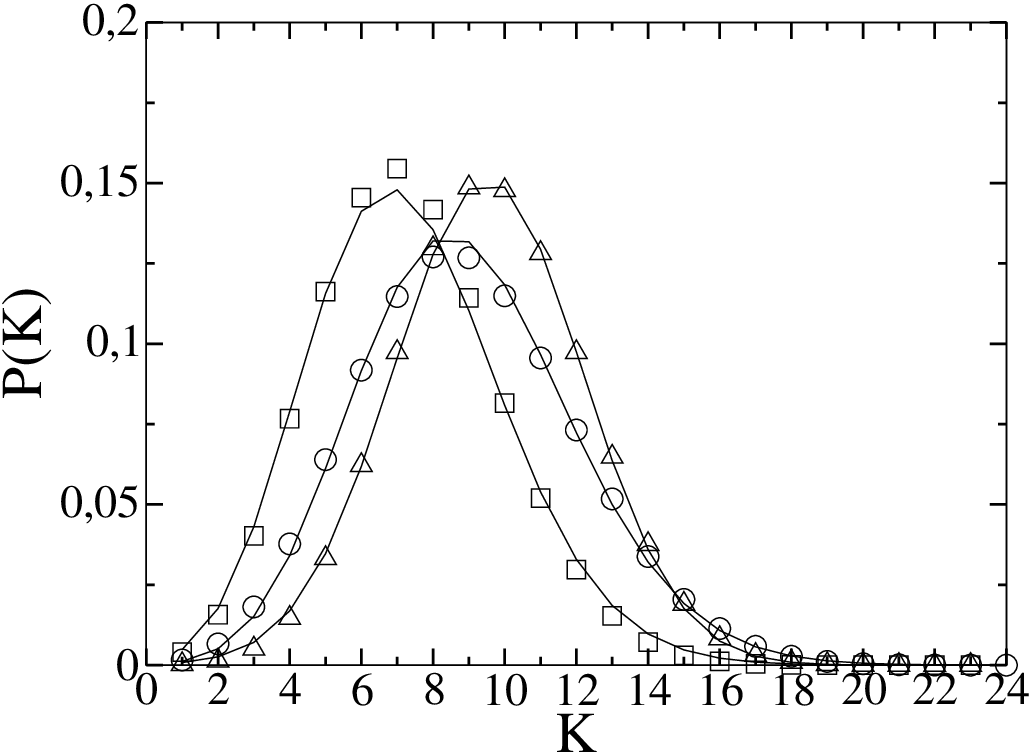}}
   \caption{The degree distribution $P(K) \times k$ of the three subnetworks in the stable $x=0.20$ and chaotic $x=0.40$ phases. (a) Stable phase: all distributions are exponential and the slopes obtained are $a$= 0.83, -1.03 and -0.95 for subnetworks $B_0$, $B_1$ and $B_2$ respectively. (b) Chaotic phase:  all  distributions are Poissonian. The different symbols refer to: ($\bigcirc$) $B=0$; ($\square$) $B=1$ and ($\triangle$) $B=2$.}
    \label{degree_all}
 \end{figure}

The degree distributions of the three subnetworks ($B_0$, $B_1$ and $B_2$) in the stable phase are finite and exponential distributions, as shown in Figure \ref{degree_all_x020}. The positive exponent for $B_0$ distribution indicates a high probability that any given node belonging to this subnetwork will be linked to almost all its neighbors, thus generating large domains of low-activity nodes on the 3-D lattice.  The negative slope obtained for $B_1$ and $B_2$ distributions indicate that in both subnetworks the active nodes have a high probability of being linked to very few neighbors. The stable regime would correspond to atypical states of the immune system with very few active populations, a feature that makes it difficult to maintain a memory regarding the perturbation history of the system, and thus explains why the mice experiments \cite{prl813034} could not be reproduced in this region of the parameter space.

As shown in Figure \ref{degree_all_x040}, in the chaotic regime the degree distributions are similar, finite and Poissonian (or bell-shaped) with an average number of connections $\langle K\rangle$ equal to $8$, $7$ and $10$ for $B_0$, $B_1$ and $B_2$ subnetworks, respectively. Therefore, theses subnetworks are random networks \cite{ErdosRenyi} as will be the entire network;  this regime, as mentioned above \cite{physA-93} would also correspond to a pathological state of the immune system, in which the order of $2/3$ of the populations are randomly activated corresponding to an overactive state equivalent to that of septicemia.

\begin{figure}[hbt]
    \centering
    \subfigure[]{\label{degree_BOBO_x030}}{\includegraphics[scale=0.33,angle=0]{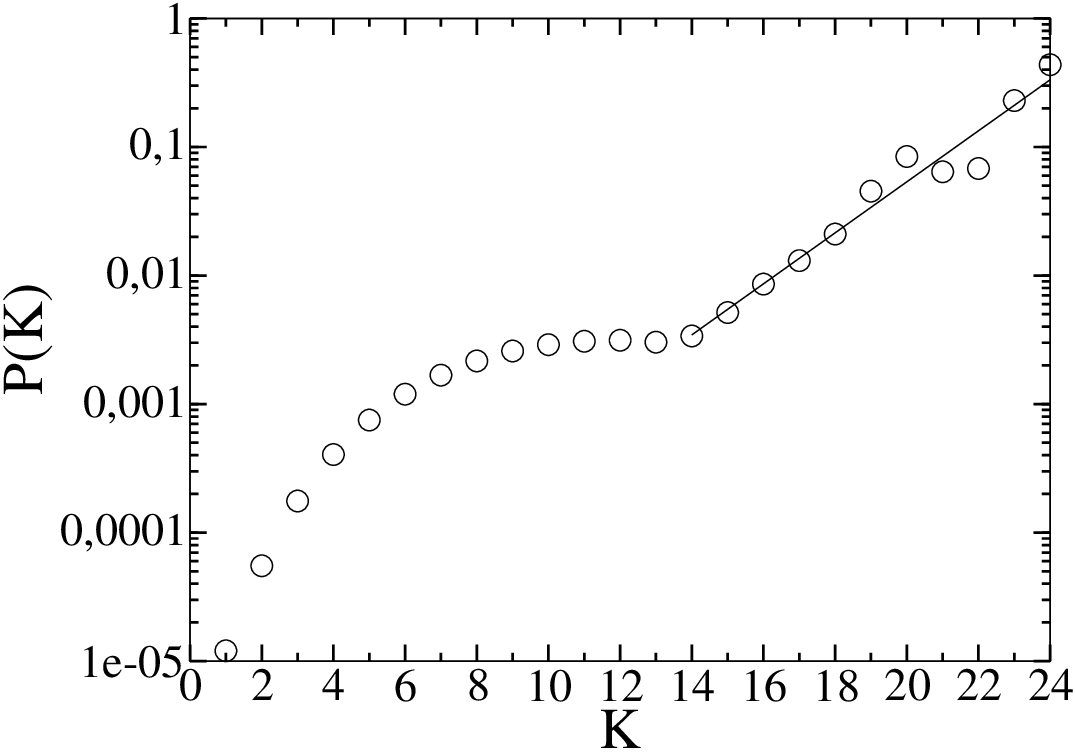}}
    \hspace*{0.25cm}\subfigure[]{\label{degree_B1B1_B2B2_x030}}{\includegraphics[scale=0.33,angle=0]{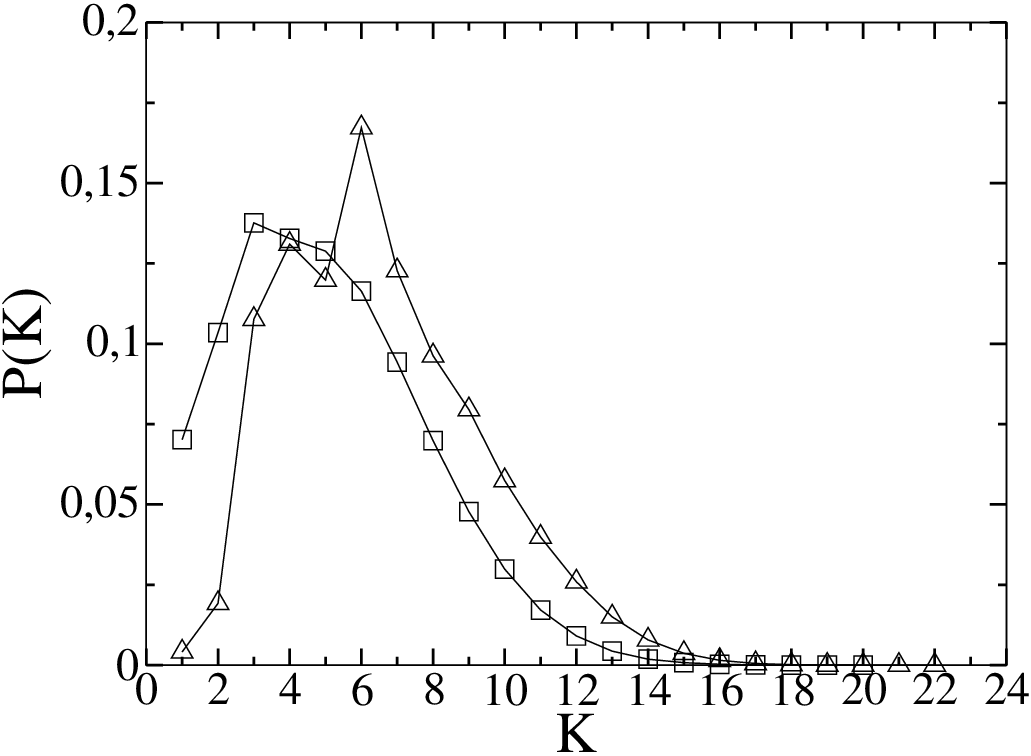}}
    \caption{Degree distribution $P(K)\times K$ at $x=0.30$. The degree distribution of subnetwork $B_0$ (a) presents mixed behavior: half exponential and half Poissonian. (b) $B_1$ and $B_2$ distributions are pure Poissonian distributions. The different symbols refer to: ($\bigcirc$) $B=0$; ($\square$) $B=1$ and ($\triangle$) $B=2$. }
    \label{degreex030}
\end{figure}

In the transition region ($x=0.30$) we find a completely different behavior of the $B_0$ distribution with respect to $B_1$ and $B_2$ distributions, as shown in Figure \ref{degreex030}. The $B_0$ degree distribution indicates mixed behavior, exhibiting characteristics of both regimes, stable and chaotic. For $K < 14$ the $B_0$ distribution is Poissonian and for $K \geq 14$ it is exponential; thus, in the transition region the $B_0$ nodes have a large probability of being linked to more than 50\% of its neighbors, forming large clusters. However, embedded in this structure there is a random network of $B_0$ nodes with low connectivity ($K<14$) exhibiting a characteristic connectivity  or an average value of $K$. As far as we know, this is the first time that such a type of mixed behavior of the degree distribution of a network has been reported on the literature. The degree distributions for $B_1$ and $B_2$ shown in Figure \ref{degree_B1B1_B2B2_x030} although showing log-normal behavior can be identified as Poisson-like distributions with long tails, corresponding to random subnetworks with few highly connected nodes (long tails). These random subnetworks have active nodes with low connectivity, and therefore, under perturbation would be able to change more easily than highly connected nodes to incorporate information into the subnetwork. The average degree for these distributions ($5$ and $6$ for $B_1$ and $B_2$ distributions respectively) is smaller than the average K obtained in the distributions of the chaotic regime, which in the case of $B_2$ is almost double.

As reported in the literature, many biological networks are scale-free and this structure would guarantee the identity and robustness of the network \cite{rmp7447,nature41141, LNP2004}. The same behavior was obtained by Burns and Ruskin \cite{PA365549} for the degree distribution of a different immune network model also based on shape-space formalism that was proposed to describe  the development of the repertoire of immune cells. The BSP model is a different model describing only B-cell population interactions based on non-local rules inspired by the functioning of the immune system. The networks it generates under certain conditions are able to reproduce the dynamical memory and behavior observed in real immune systems. Poissonian behavior predominates in the degree distribution for the subnetworks in transition and the chaotic phases. These results are very interesting since they suggest that random structures are always present in the active subnetworks of these regimes and would be responsible for the dynamic allocation of memory observed in the transition region. The randomly attached active sites of low connectivity allow for rapid changes to add or discard information. The plasticity, which permits  the memory of previous antigen (perturbation) history to be maintained, only observed on the transition region, will also depend on the behavior of the degree distribution of low concentration subnetworks exhibiting a mixed behavior (exponential and pure Poissonian) \ref{degreex030}. In order to achieve the plasticity necessary to maintain the history of the perturbation (antigens) memory,  it is necessary for the low connectivity subnetwork to have a random structure embedded on one that is very connected and almost regular: low connectivity sites would allow for rapid changes following the active populations while the high connected ones would keep the robustness and the integrity of the system.  In the chaotic regime the fact that all distributions are Poissonian with a relatively low average connectivity allows for rapid changes in all subnetworks, and since the densities of different populations are of the same order of magnitude rapid changes would very easily lead to  the over-activation of the populations. Therefore, the randomness on the degree distribution is an important feature for the immune network \cite {Jerne} to maintain a memory of the previous history of antigen presentations. The exponential behavior observed on the degree distribution for all subnetworks in the stable regime do not allow the memory to be maintained for two reasons:  overall, the high density of low concentration sites is highly connected, which is an unfavorable aspect regarding the necessary changes to incorporate information, and the very low density of active sites does not allow for the accumulation of information.

\begin{figure}[hbt]
   \centering
   \subfigure[]{\label{knn_x20}}{\includegraphics[scale=0.33,angle=0]{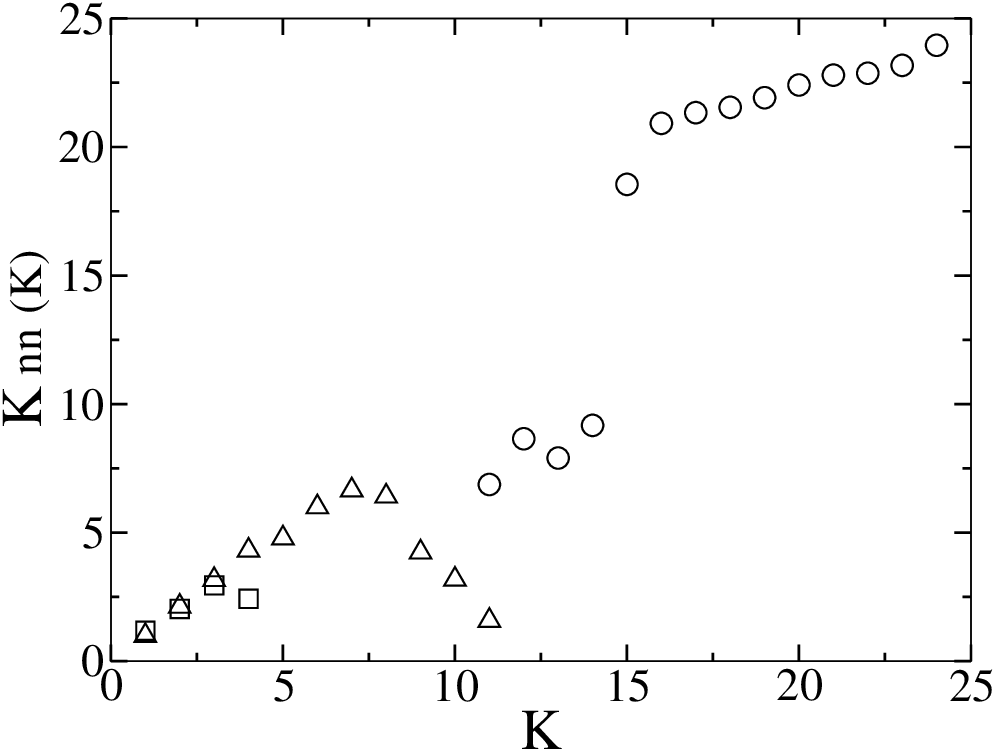}}
    \hspace*{0.5cm}\subfigure[]{\label{knn_x30}}{\includegraphics[scale=0.33,angle=0]{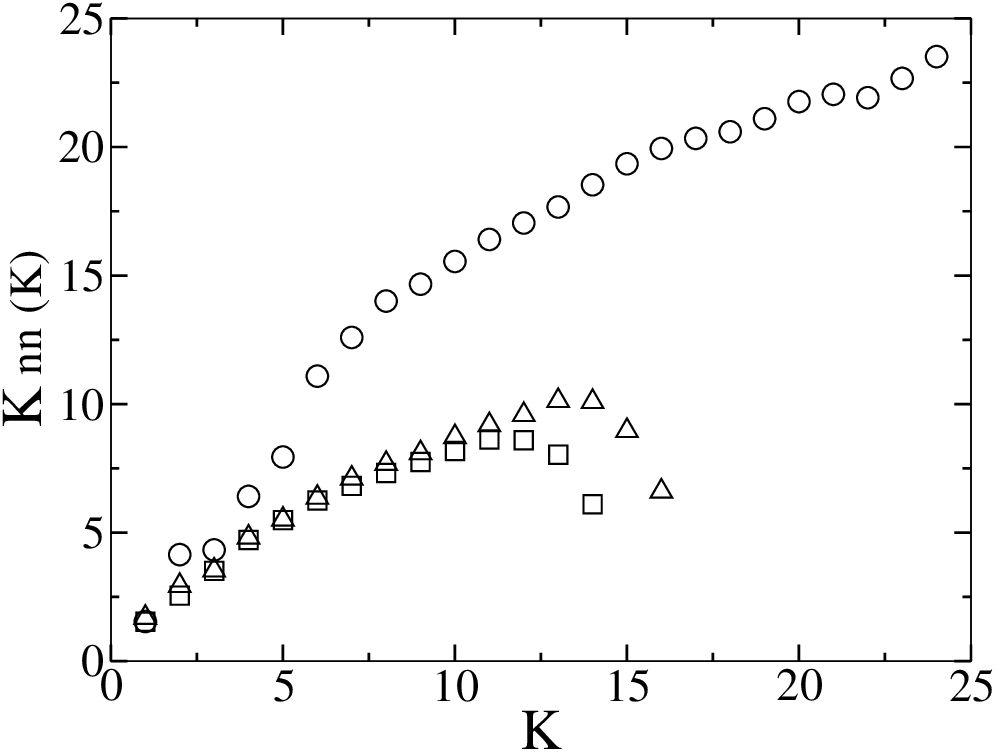}}\\
    \hspace*{0.5cm}\subfigure[]{\label{knn_x40}}{\includegraphics[scale=0.33,angle=0]{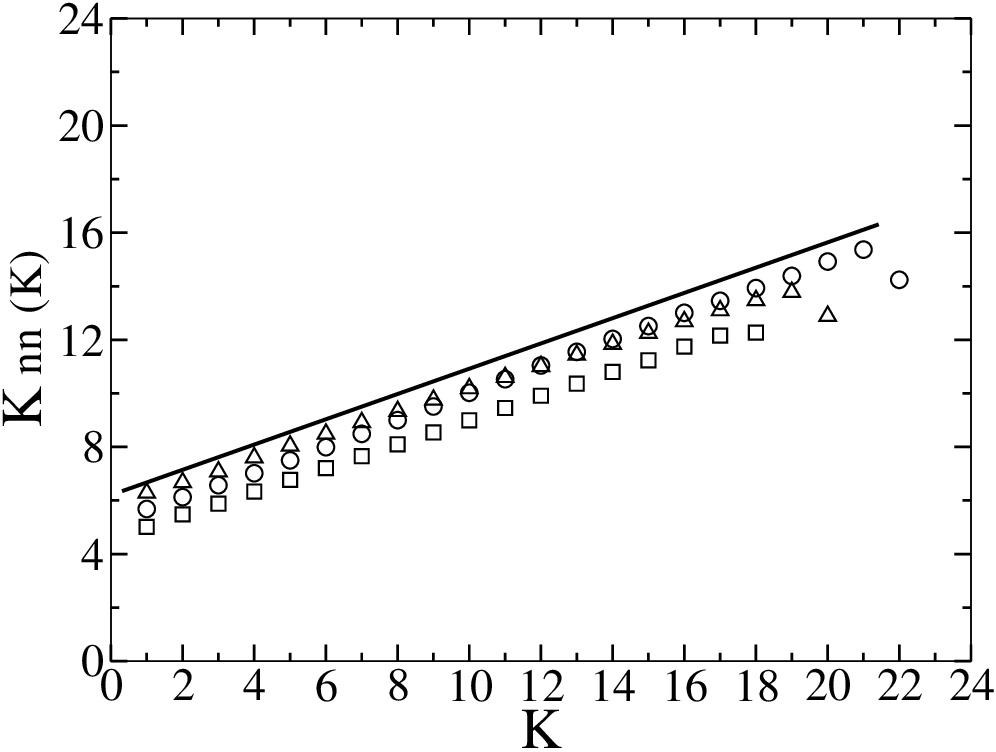}}\\
   \caption{Average degree of the nearest neighbors $K_{nn}(K)$ for nodes with degree $K$. An assortative mixed behavior is observed for sub networks ($\bigcirc$) $B_0$; ($\square$)  $B_1$ and ($\triangle$ $B_2$) at: (a) Stable phase ($x=0.20$),  (b) Transition region ($x=0.30$)  and (c) the chaotic region ($x=0.40$), for which the slope of the solid guide line is 0.44.}
   \label{knn}
\end{figure}

We have also studied the behavior of the average degree $K_{nn}(K)$ of the neighbors of the nodes with degree $K$. Overall, the results indicate that there is a positive correlation between the neighbors or assortative mixing properties, i.e., nodes with the same type of connectivity (low and high) tend to cluster together. Moreover, results indicate that there is an overall linear growth of $K_{nn}(K)$ as a function of K, but behavior is different for all three phases. In the stable regime $K_{nn}(K)$ grows almost linearly (positive correlation) for $K \geq 14$, $K\leq 3$ and $K\leq 7$ for $B_0$, $B_1$ and $B-2$ distributions, respectively, but decreasing very rapidly (negative correlation) afterwards in the last two distributions (Fig. \ref{knn_x20}). In other words the positive correlation grows up to a critical value ($K_c$) of connectivity and becomes negative for $K \geq K_c$.  At the transition we observed that the average degree for $B_0$ distribution behaves linearly for all K, although with different slopes in at least two regions; the behavior of $B_1$ and $B_2$ persists as in the ordered phase, but in this case they grow linearly for $K\leq 12$ and $K\leq 14$ respectively, with a subsequent decrease, as in the previous regime (Fig. \ref{knn_x30}). The existence of cut-off degrees is characteristic of finite random networks, however the linear growth has not previously been observed. In the chaotic region ($x=0.40$) $K_{nn}(K)$ grows linearly with $K$ ((see Figure \ref{knn_x40}) for all subnetworks, as expected, since in this region, the degree distribution is Poissonian and the densities of nodes for all subnetworks are of the same magnitude. The positive correlation would allow for the over-activation of nodes as observed. 

For the same set of parameters $N=28^3$, in the transition and chaotic regions ($x=0.3$ and $x=0.4$ respectively) we have estimated the mean shortest path $L$ for the three subnetworks.  Due to the computational cost it is difficult to perform this calculation with the necessary statistics, which is why here we present the results obtained for a simple sampling. At the transition region we have obtained   $L=10.72$ $(N=14433)$ for $B_0$, $L=13.05$ $(N=2872)$ for $B_1$ and $L=11.60$ $(N=4338)$ for $B_2$ and for the chaotic regime ($x=0.40$) the results are:  $L=11.38$ $(N=7471)$ for $B_0$, $L=11.76$ $(N=6289)$ for $B_1$ and $L=11.19$ $(N=8185)$ for $B_2$. In all cases we observe that $L\ll N$ (number of nodes in the subnetwork) but overall $log N \leq L \leq L^{1/d}$, i.e. the behavior between regular lattice and random networks. Therefore, $L \ll N$ is not necessarily a signature of small-world properties as suggested by other biological networks \cite{nature393440}, but in our case it is a signature of the existence of very connected nodes. We should expect small-word properties as in other random networks, however due to the computational cost we have not looked for evidence of such properties.

\begin{figure}[hbt]
    \centering
    \includegraphics[scale=0.33,angle=0]{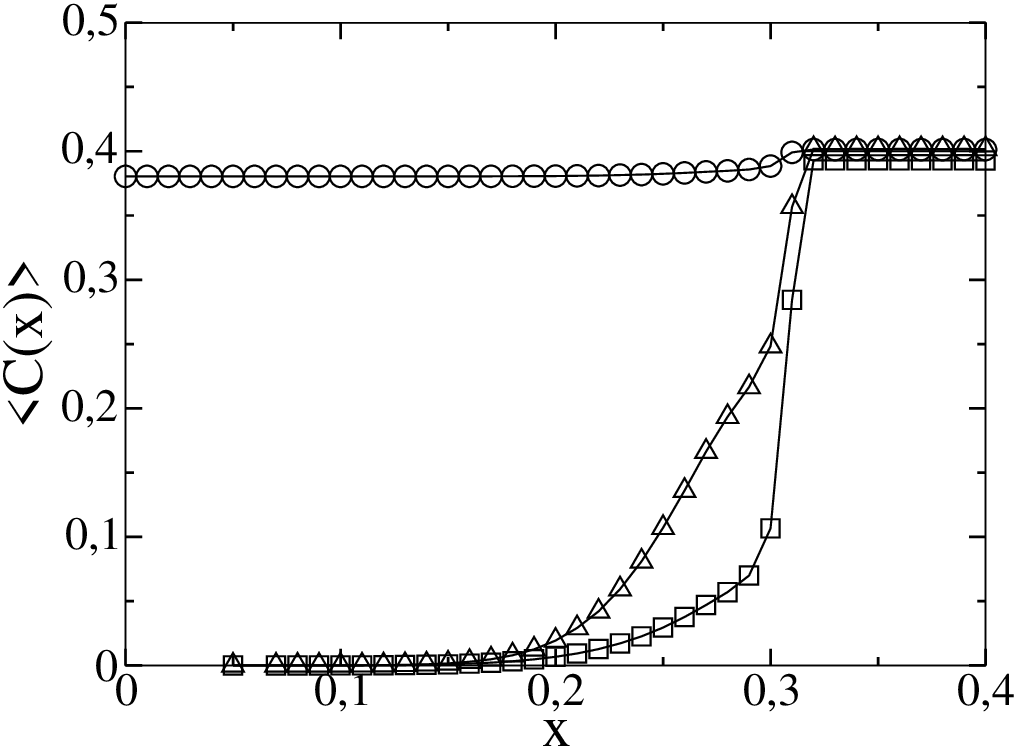}
    \caption{The global clustering coefficient $C$ as a function of $x$ for the subnetworks: ($\bigcirc$)  $B_0$; ($\square$)$B_1$ and ($\triangle$)$B_2$.}
    \label{Cx_all}
\end{figure}

When we compute the clustering coefficient $C$ as a function of $x$ for the three subnetworks (shown in Figure \ref{Cx_all}), we observe that for the subnetwork $B_0$ $C$ is approximately constant, with a slight increase from $0.38$ to $0.40$ in the transition region. This result indicates a strong aggregation between the populations at a low concentration, i.e., on average around $40\%$ percent of the neighbors are neighbors among themselves.  However, the behavior of the clustering coefficients for subnetworks $B_1$ and $B_2$ differ completely from that of $B_0$: for small $x$ in both cases the clustering coefficient is close to zero and start to increase around  $x=0.20$, reaching its maximum value ($\sim 0.4$) for $x>0.3$. The increase of the clustering coefficient implies an increase in the aggregation of these subnetworks, reflecting an increase in the concentrations of active sites with intermediate and high concentrations as $x$ increases (Figure \ref {density}). We have also observed that the cohesiveness between different populations is always less significant than the cohesiveness among the elements of the same population (results not shown).

\begin{figure} [hbt]
  \centering
   \subfigure[]{\label{CK_B1B1_B2B2_x020}}{\includegraphics[scale=0.33,angle=0]{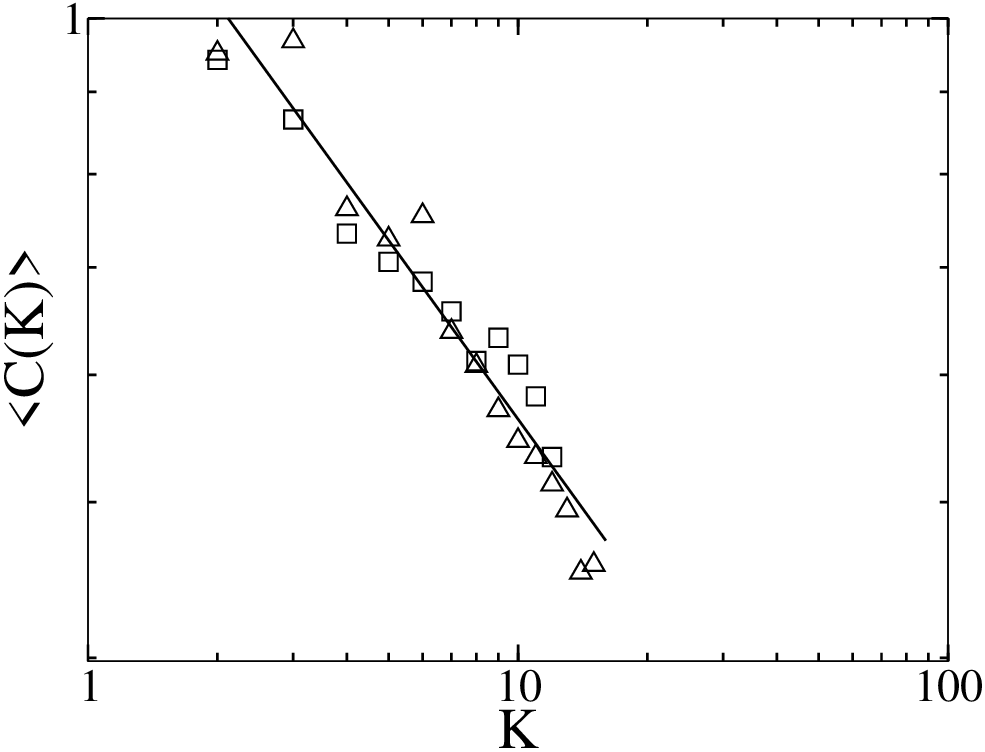}}
   \hspace*{0.25cm}\subfigure[]{\label{CK_all_x040}}{\includegraphics[scale=0.33,angle=0]{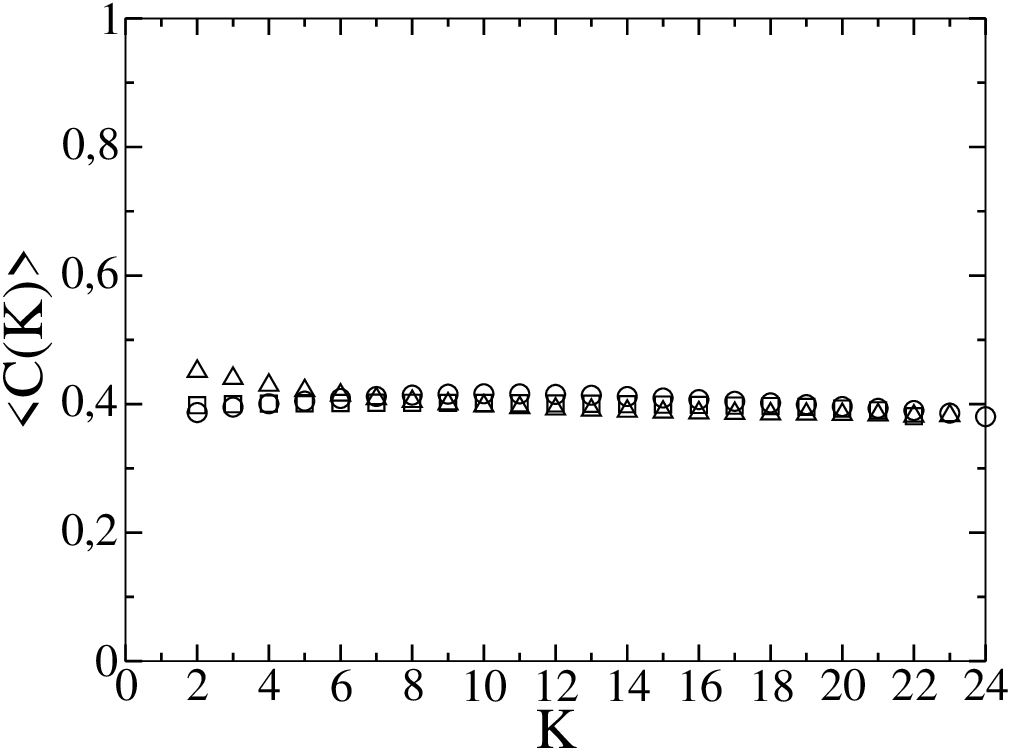}}
   \caption{Average clustering coefficient $\langle C(K)\rangle$ as a function of the connectivity $K$ for (a) Stable ($x=0.20$) and (b) Chaotic  ($x=0.40$) phases, where ($\bigcirc$) corresponds to $B_0$, ($\square$) to $B_1$ and ($\triangle$)  $B_2$ sub networks, respectively.}
   \label{CK}
\end{figure}

While the average clustering coefficient is almost constant as a function of $K$ in the chaotic and transition regions, it exhibits a power-law behavior in the stable region ($x=0.2$), as shown in Figure \ref{CK_B1B1_B2B2_x020}. This behavior indicates that the subnetworks in the stable regime have hierarchical structures \cite{pre67026112} with exponents that are smaller than unity ($0.29$ for $B_1$ and $0.40$ for $B_2$) as observed in the case of the internet \cite {VPSV}. Hierarchical structures go beyond simple clustering by including organization at all scales in the network simultaneously. Trees, in which the nodes are connected, generally possess common ancestors that are topologically closer to them than nodes, which are not closely related. Since the numbers of populations with intermediate or high concentration are small in the stable regime, this would explain why under perturbation the system changes, incorporating information regarding perturbation and losing information on the previous perturbation, do not maintain any memory regarding its history \cite{epjb34119, Zorzenon-Copelli03}.

Figure \ref{CK_all_x040} shows the constant behavior of the average clustering coefficient as a function of K in the chaotic region. Constant behavior is characteristic of random networks where the clustering coefficient is $C_i=p$ where $p=\langle k\rangle/N$ is the probability of connection between two nodes. For random networks the small aggregation is due to small  $p$  \cite{ErdosRenyi}. Previous results indicate that in the chaotic region the subnetworks are random networks.

\section{Conclusions}
\label{discussion}

In this work, from the viewpoint of a complex network we have analyzed an immune network model, which was able to reproduce the behavior of the immune systems of mice. Indeed, the immune network is complex, and contrary to what has been reported in the literature regarding other biological networks, this biological network is characterized by random behavior rather than scale-free \cite{rmp7447}.

In fact, the networks obtained in the three dynamical regimes observed in the model are composed of three different subnetworks of low, intermediate and high concentrations.  In this work we have investigated the behavior of the three subnetworks in the different regimes.  We observed the existence of random characteristics in the three regions:  in the ordered phase the degree distributions are purely exponential; in the transition region while $B_0$ mix exponential ($K \leq 14$) and Poisson ($K>14$), the $B_1$ and $B_2$ are Possonian; and in the chaotic region the distributions are purely Poissonian.  In reality, the immune network seems regulated by random subnetworks of active populations ($B_1$ and $B_2$). The randomness of these structures formed by nodes with low average connectivity (with respect to the maximum number of neighbors) allow for rapid changes that add or discard network information. However the necessary plasticity that would allow for adaptation and maintainance of  the network’s  identity and memory regarding its history in terms of perturbations (or antigen presentations) depends on the structure of the low concentration population subnetwork. The structure that allows the reproduction of  immune system behavior in mice is composed of random substructures embedded in one that is very connected: low concentration sites with random low connectivity will allow for the incorporation of new information (new active populations in the network) leading to a different configuration that when compared to the previous one has added relevant new information and discarded  the unnecessary.

Another important result that differs from those reported in the literature, is the fact that $K_{nn}(K)$ grows linearly as a function of K. For stable and transition regions the active population subnetworks show a linear growth (positive correlation) with a cut-off $K_c$, as expected for finite random networks, followed by a negative correlation.  However,  $B_0$ distributions do not have a cut-off; in the chaotic region, all distributions exhibit a linear growth of $K_{nn}(K)$ as a function of K. Put another way, while the $B_0$ subnetworks exhibit assortative mixing in all phases, there is an interplay between assortative and disassortative mixing  in $B_1$ and $B_2$ subnetworks at stable and transition regions, depending on the value of $K$. If $K \leq K_c$  there is a positive, linear degree-correlation, but for $K >K_c$ the correlation is negative. There is a maximum threshold value in the transition region but that disappears in the chaotic region giving way to pure assortative behavior. This suggests that although random, the networks obtained with this model are much more complex than those  thus far reported in the literature. We believe that the differences between our results and those  from the literature arise from the  fact that in this study the network´s  topology reflects the dynamics inspired in the biological process attributed to the immune system, while the majority of biological networks that have been studied neither include nor reflect  the dynamics of the process, since in many cases it is unknown.

\section*{Acknowledgments}

We thank Silvio Ferreira Jr. for his helpful discussions and suggestions. This study received financial support from the following Brazilian institutions: the Conselho Nacional de Desenvolvimento Cient\'ifico e Tecnol\'ogico (CNPq, http://www.cnpq.br), the Coordena\c c\~ao de Aperfeioamento de Pessoal de Nivel Superior (CAPES, http://www.capes.gov.br) and the Funda\c c\~ao de Amparo Ci\^encia e Tecnologia do Estado de Pernambuco (FACEPE, http://www.facepe.br-Pronex-EDT 0012-05.03/04 and Pronex-APQ 0203-1.05/08).

\bibliography{bibli}

\begin{thebibliography}{10}%
\makeatletter
\providecommand \@ifxundefined [1]{%
 \ifx #1\undefined \expandafter \@firstoftwo
 \else \expandafter \@secondoftwo
\fi
}%
\providecommand \@ifnum [1]{%
 \ifnum #1\expandafter \@firstoftwo
 \else \expandafter \@secondoftwo
\fi
}%
\providecommand \enquote [1]{``#1''}%
\providecommand \bibnamefont  [1]{#1}%
\providecommand \bibfnamefont [1]{#1}%
\providecommand \citenamefont [1]{#1}%
\providecommand\href[0]{\@sanitize\@href}%
\providecommand\@href[1]{\endgroup\@@startlink{#1}\endgroup\@@href}%
\providecommand\@@href[1]{#1\@@endlink}%
\providecommand \@sanitize [0]{\begingroup\catcode`\&12\catcode`\#12\relax}%
\@ifxundefined \pdfoutput {\@firstoftwo}{%
 \@ifnum{\z@=\pdfoutput}{\@firstoftwo}{\@secondoftwo}%
}{%
 \providecommand\@@startlink[1]{\leavevmode\special{html:<a href="#1">}}%
 \providecommand\@@endlink[0]{\special{html:</a>}}%
}{%
 \providecommand\@@startlink[1]{%
  \leavevmode
  \pdfstartlink
   attr{/Border[0 0 1 ]/H/I/C[0 1 1]}%
   user{/Subtype/Link/A<</Type/Action/S/URI/URI(#1)>>}%
  \relax
 }%
 \providecommand\@@endlink[0]{\pdfendlink}%
}%
\providecommand \url  [0]{\begingroup\@sanitize \@url }%
\providecommand \@url [1]{\endgroup\@href {#1}{\urlprefix}}%
\providecommand \urlprefix [0]{URL }%
\providecommand \Eprint[0]{\href }%
\@ifxundefined \urlstyle {%
  \providecommand \doi [1]{doi:\discretionary{}{}{}#1}%
}{%
  \providecommand \doi [0]{doi:\discretionary{}{}{}\begingroup
  \urlstyle{rm}\Url }%
}%
\providecommand \doibase [0]{http://dx.doi.org/}%
\providecommand \Doi[1]{\href{\doibase#1}}%
\providecommand \bibAnnote [3]{%
  \BibitemShut{#1}%
  \begin{quotation}\noindent
    \textsc{Key:}\ #2\\\textsc{Annotation:}\ #3%
  \end{quotation}%
}%
\providecommand \bibAnnoteFile [2]{%
  \IfFileExists{#2}{\bibAnnote {#1} {#2} {\input{#2}}}{}%
}%
\providecommand \typeout [0]{\immediate \write \m@ne }%
\providecommand \selectlanguage [0]{\@gobble}%
\providecommand \bibinfo [0]{\@secondoftwo}%
\providecommand \bibfield [0]{\@secondoftwo}%
\providecommand \translation [1]{[#1]}%
\providecommand \BibitemOpen[0]{}%
\providecommand \bibitemStop [0]{}%
\providecommand \bibitemNoStop [0]{.\EOS\space}%
\providecommand \EOS [0]{\spacefactor3000\relax}%
\providecommand \BibitemShut [1]{\csname bibitem#1\endcsname}%
\bibitem{Janeway}%
  \BibitemOpen
  \bibfield{author}{%
  \bibinfo {author} {\bibfnamefont{C.~A.}\ \bibnamefont{Janeway}}, \bibinfo
  {author} {\bibfnamefont{P.~A.}\ \bibnamefont{Traver}}, \bibinfo {author}
  {\bibfnamefont{M.}~\bibnamefont{Walport}},\ and\ \bibinfo {author}
  {\bibfnamefont{J.}~\bibnamefont{Capra}},\ }%
  \emph{\bibinfo {title} {Immunobiology: The Immune System In Health And
  Disease}}\ (\bibinfo {publisher} {Garland Science Publishing},\ \bibinfo
  {address} {NY},\ \bibinfo {year} {2009})%
  \bibAnnoteFile{NoStop}{Janeway}%
\bibitem{Jerne}%
  \BibitemOpen
  \bibfield{author}{%
  \bibinfo {author} {\bibfnamefont{N.~K.}\ \bibnamefont{Jerne}},\ }%
  \bibfield{journal}{%
  \bibinfo {journal} {Ann. Immuno. (Inst. Pasteur)}\ }%
  \textbf{\bibinfo {volume} {125 C}},\ \bibinfo {pages} {372} (\bibinfo {year}
  {1974})%
  \bibAnnoteFile{NoStop}{Jerne}%
\bibitem{Coutinho89}%
  \BibitemOpen
  \bibfield{author}{%
  \bibinfo {author} {\bibfnamefont{A.}~\bibnamefont{Coutinho}},\ }%
  \bibfield{journal}{%
  \bibinfo {journal} {Immunol. Rev.}\ }%
  \textbf{\bibinfo {volume} {110}},\ \bibinfo {pages} {63} (\bibinfo {year}
  {1989})%
  \bibAnnoteFile{NoStop}{Coutinho89}%
\bibitem{Holmberg-Forsgren}%
  \BibitemOpen
  \bibfield{author}{%
  \bibinfo {author} {\bibfnamefont{D.}~\bibnamefont{Holmberg}}, \bibinfo
  {author} {\bibnamefont{Anderson}}, \bibinfo {author}
  {\bibfnamefont{L.}~\bibnamefont{Carlsson}},\ and\ \bibinfo {author}
  {\bibfnamefont{S.}~\bibnamefont{Forsgren}},\ }%
  \bibfield{journal}{%
  \bibinfo {journal} {Immunol. Rev.}\ }%
  \textbf{\bibinfo {volume} {110}},\ \bibinfo {pages} {89} (\bibinfo {year}
  {1989})%
  \bibAnnoteFile{NoStop}{Holmberg-Forsgren}%
\bibitem{Lundkvist-Holmberg}%
  \BibitemOpen
  \bibfield{author}{%
  \bibinfo {author} {\bibfnamefont{I.}~\bibnamefont{Lundkvist}}, \bibinfo
  {author} {\bibfnamefont{A.}~\bibnamefont{Coutinho}}, \bibinfo {author}
  {\bibfnamefont{F.}~\bibnamefont{Varela}},\ and\ \bibinfo {author}
  {\bibfnamefont{D.}~\bibnamefont{Holmberg}},\ }%
  \bibfield{journal}{%
  \bibinfo {journal} {Proc. Natl. Acad. Sci. USA}\ }%
  \textbf{\bibinfo {volume} {86}},\ \bibinfo {pages} {5074} (\bibinfo {year}
  {1989})%
  \bibAnnoteFile{NoStop}{Lundkvist-Holmberg}%
\bibitem{physA18042}%
  \BibitemOpen
  \bibfield{author}{%
  \bibinfo {author} {\bibfnamefont{D.}~\bibnamefont{Stauffer}}\ and\ \bibinfo
  {author} {\bibfnamefont{G.}~\bibnamefont{Weisbuch}},\ }%
  \bibfield{journal}{%
  \bibinfo {journal} {Physica A}\ }%
  \textbf{\bibinfo {volume} {180}},\ \bibinfo {pages} {42} (\bibinfo {year}
  {1992})%
  \bibAnnoteFile{NoStop}{physA18042}%
\bibitem{DeBoer-Segel-Perelson}%
  \BibitemOpen
  \bibfield{author}{%
  \bibinfo {author} {\bibfnamefont{R.~J.~D.}\ \bibnamefont{Boer}}, \bibinfo
  {author} {\bibfnamefont{L.~A.}\ \bibnamefont{Segel}},\ and\ \bibinfo {author}
  {\bibfnamefont{A.~S.}\ \bibnamefont{Perelson}},\ }%
  \bibfield{journal}{%
  \bibinfo {journal} {J. Theor. Biol.}\ }%
  \textbf{\bibinfo {volume} {155}},\ \bibinfo {pages} {295} (\bibinfo {year}
  {1992})%
  \bibAnnoteFile{NoStop}{DeBoer-Segel-Perelson}%
\bibitem{physa2191}%
  \BibitemOpen
  \bibfield{author}{%
  \bibinfo {author} {\bibfnamefont{R.~M.~Z.}\ \bibnamefont{dos Santos}}\ and\
  \bibinfo {author} {\bibfnamefont{A.~T.}\ \bibnamefont{Bernades}},\ }%
  \bibfield{journal}{%
  \bibinfo {journal} {Physica A}\ }%
  \textbf{\bibinfo {volume} {219}},\ \bibinfo {pages} {1} (\bibinfo {year}
  {1995})%
  \bibAnnoteFile{NoStop}{physa2191}%
\bibitem{physA-93}%
  \BibitemOpen
  \bibfield{author}{%
  \bibinfo {author} {\bibfnamefont{R.~M.~Z.}\ \bibnamefont{dos Santos}},\ }%
  \bibfield{journal}{%
  \bibinfo {journal} {Physica A}\ }%
  \textbf{\bibinfo {volume} {196}},\ \bibinfo {pages} {12} (\bibinfo {year}
  {1993})%
  \bibAnnoteFile{NoStop}{physA-93}%
\bibitem{jtb186173}%
  \BibitemOpen
  \bibfield{author}{%
  \bibinfo {author} {\bibfnamefont{A.~T.}\ \bibnamefont{Bernardes}}\ and\
  \bibinfo {author} {\bibfnamefont{R.~M.~Z.}\ \bibnamefont{{dos Santos}}},\ }%
  \bibfield{journal}{%
  \bibinfo {journal} {J. Theor. Biol.}\ }%
  \textbf{\bibinfo {volume} {186}},\ \bibinfo {pages} {173} (\bibinfo {year}
  {1997})%
  \bibAnnoteFile{NoStop}{jtb186173}%
\bibitem{prl813034}%
  \BibitemOpen
  \bibfield{author}{%
  \bibinfo {author} {\bibfnamefont{R.~M.~Z.}\ \bibnamefont{dos Santos}}\ and\
  \bibinfo {author} {\bibfnamefont{A.~T.}\ \bibnamefont{Bernades}},\ }%
  \bibfield{journal}{%
  \bibinfo {journal} {Phys. Rev. Let.}\ }%
  \textbf{\bibinfo {volume} {81}},\ \bibinfo {pages} {3034} (\bibinfo {year}
  {1998})%
  \bibAnnoteFile{NoStop}{prl813034}%
\bibitem{Zorzenon-Copelli03}%
  \BibitemOpen
  \bibfield{author}{%
  \bibinfo {author} {\bibfnamefont{R.~M.~Z.}\ \bibnamefont{dos Santos}}\ and\
  \bibinfo {author} {\bibfnamefont{M.}~\bibnamefont{Copelli}},\ }%
  \bibfield{journal}{%
  \bibinfo {journal} {Brazilian Journal of Physics}\ }%
  \textbf{\bibinfo {volume} {33}},\ \bibinfo {pages} {628} (\bibinfo {year}
  {2003})%
  \bibAnnoteFile{NoStop}{Zorzenon-Copelli03}%
\bibitem{epjb34119}%
  \BibitemOpen
  \bibfield{author}{%
  \bibinfo {author} {\bibfnamefont{M.}~\bibnamefont{Copelli}}, \bibinfo
  {author} {\bibfnamefont{R.~M.~Z.}\ \bibnamefont{dos Santos}},\ and\ \bibinfo
  {author} {\bibfnamefont{D.~A.}\ \bibnamefont{Stariolo}},\ }%
  \bibfield{journal}{%
  \bibinfo {journal} {Eur. Phys. J. B}\ }%
  \textbf{\bibinfo {volume} {34}},\ \bibinfo {pages} {119} (\bibinfo {year}
  {2003})%
  \bibAnnoteFile{NoStop}{epjb34119}%
\bibitem{rmp7447}%
  \BibitemOpen
  \bibfield{author}{%
  \bibinfo {author} {\bibfnamefont{R.}~\bibnamefont{Albert}}\ and\ \bibinfo
  {author} {\bibfnamefont{A.-L.}\ \bibnamefont{Barab{\'a}si}},\ }%
  \bibfield{journal}{%
  \bibinfo {journal} {Rev. Mod. Phys.}\ }%
  \textbf{\bibinfo {volume} {74}},\ \bibinfo {pages} {47} (\bibinfo {year}
  {2002})%
  \bibAnnoteFile{NoStop}{rmp7447}%
\bibitem{nature41141}%
  \BibitemOpen
  \bibfield{author}{%
  \bibinfo {author} {\bibfnamefont{H.}~\bibnamefont{Jeong}}, \bibinfo {author}
  {\bibfnamefont{S.~P.}\ \bibnamefont{Mason}}, \bibinfo {author}
  {\bibfnamefont{Z.~N.}\ \bibnamefont{Oltvai}},\ and\ \bibinfo {author}
  {\bibfnamefont{A.~L.}\ \bibnamefont{Barab{\'a}si}},\ }%
  \bibfield{journal}{%
  \bibinfo {journal} {Nature}\ }%
  \textbf{\bibinfo {volume} {411}},\ \bibinfo {pages} {41} (\bibinfo {year}
  {2001})%
  \bibAnnoteFile{NoStop}{nature41141}%
\bibitem{jtb81645}%
  \BibitemOpen
  \bibfield{author}{%
  \bibinfo {author} {\bibfnamefont{A.~S.}\ \bibnamefont{Perelson}}\ and\
  \bibinfo {author} {\bibfnamefont{G.~F.}\ \bibnamefont{Oster}},\ }%
  \bibfield{journal}{%
  \bibinfo {journal} {J. Theor. Biol.}\ }%
  \textbf{\bibinfo {volume} {81}},\ \bibinfo {pages} {645} (\bibinfo {year}
  {1979})%
  \bibAnnoteFile{NoStop}{jtb81645}%
\bibitem{Bernardes-Zorzenon01}%
  \BibitemOpen
  \bibfield{author}{%
  \bibinfo {author} {\bibfnamefont{A.~T.}\ \bibnamefont{Bernardes}}\ and\
  \bibinfo {author} {\bibfnamefont{R.~M.~Z.}\ \bibnamefont{dos Santos}},\ }%
  \bibfield{journal}{%
  \bibinfo {journal} {Int. J. Mod. Phys. C}\ }%
  \textbf{\bibinfo {volume} {12}},\ \bibinfo {pages} {1} (\bibinfo {year}
  {2001})%
  \bibAnnoteFile{NoStop}{Bernardes-Zorzenon01}%
\bibitem{pre67026112}%
  \BibitemOpen
  \bibfield{author}{%
  \bibinfo {author} {\bibfnamefont{E.}~\bibnamefont{Ravasz}}\ and\ \bibinfo
  {author} {\bibfnamefont{A.-L.}\ \bibnamefont{Barab{\'a}si}},\ }%
  \bibfield{journal}{%
  \bibinfo {journal} {Phys. Rev. E}\ }%
  \textbf{\bibinfo {volume} {67}} (\bibinfo {year} {2003})%
  \bibAnnoteFile{NoStop}{pre67026112}%
\bibitem{knn_book}%
  \BibitemOpen
  \bibfield{author}{%
  \bibinfo {author} {\bibfnamefont{A.}~\bibnamefont{Barrat}}, \bibinfo {author}
  {\bibfnamefont{M.}~\bibnamefont{Barth\'elemy}},\ and\ \bibinfo {author}
  {\bibfnamefont{A.}~\bibnamefont{Vespignani}},\ }%
  \emph{\bibinfo {title} {Dynamical Processes on Complex Networks}}\ (\bibinfo
  {publisher} {Cambridge Univ. Press},\ \bibinfo {year} {2009})%
  \bibAnnoteFile{NoStop}{knn_book}%
\bibitem{ErdosRenyi}%
  \BibitemOpen
  \bibfield{author}{%
  \bibinfo {author} {\bibfnamefont{P.}~\bibnamefont{Erd\H{o}s}}\ and\ \bibinfo
  {author} {\bibfnamefont{A.}~\bibnamefont{R{\'e}nyi}},\ }%
  \bibfield{journal}{%
  \bibinfo {journal} {Publ. Math.}\ }%
  \textbf{\bibinfo {volume} {6}} (\bibinfo {year} {1959})%
  \bibAnnoteFile{NoStop}{ErdosRenyi}%
\bibitem{LNP2004}%
  \BibitemOpen
  \bibfield{author}{%
  \bibinfo {author} {\bibfnamefont{B.}~\bibnamefont{A-L}}, \bibinfo {author}
  {\bibfnamefont{Z.~N.}\ \bibnamefont{Oltvai}},\ and\ \bibinfo {author}
  {\bibfnamefont{S.}~\bibnamefont{Wuchty}},\ }%
  \bibfield{journal}{%
  \bibinfo {journal} {Lect. Notes. Phys.}\ }%
  \textbf{\bibinfo {volume} {650}},\ \bibinfo {pages} {443} (\bibinfo {year}
  {2004})%
  \bibAnnoteFile{NoStop}{LNP2004}%
\bibitem{PA365549}%
  \BibitemOpen
  \bibfield{author}{%
  \bibinfo {author} {\bibfnamefont{H.~J.}\ \bibnamefont{Ruskin}}\ and\ \bibinfo
  {author} {\bibfnamefont{J.}~\bibnamefont{Burns}},\ }%
  \bibfield{journal}{%
  \bibinfo {journal} {Physica A}\ }%
  \textbf{\bibinfo {volume} {365}},\ \bibinfo {pages} {549} (\bibinfo {year}
  {2006})%
  \bibAnnoteFile{NoStop}{PA365549}%
\bibitem{nature393440}%
  \BibitemOpen
  \bibfield{author}{%
  \bibinfo {author} {\bibfnamefont{D.~J.}\ \bibnamefont{Watts}}\ and\ \bibinfo
  {author} {\bibfnamefont{S.~H.}\ \bibnamefont{Strogatz}},\ }%
  \bibfield{journal}{%
  \bibinfo {journal} {Nature}\ }%
  \textbf{\bibinfo {volume} {393}},\ \bibinfo {pages} {440} (\bibinfo {year}
  {1998})%
  \bibAnnoteFile{NoStop}{nature393440}%
\bibitem{VPSV}%
  \BibitemOpen
  \bibfield{author}{%
  \bibinfo {author} {\bibfnamefont{A.}~\bibnamefont{V\'azquez}}, \bibinfo
  {author} {\bibfnamefont{R.}~\bibnamefont{Pastor-Satorras}},\ and\ \bibinfo
  {author} {\bibfnamefont{A.}~\bibnamefont{Vespignani}},\ }%
  \bibfield{journal}{%
  \bibinfo {journal} {Phys. Rev. E}\ }%
  \textbf{\bibinfo {volume} {65}},\ \bibinfo {pages} {066130} (\bibinfo {year}
  {2002})%
  \bibAnnoteFile{NoStop}{VPSV}%
\end{thebibliography}%

\end{document}